\documentclass[final,3p,times]{elsarticle}

\usepackage{graphicx}

\usepackage{amssymb}

\usepackage[english,french]{babel}

\def\og{\leavevmode\raise.3ex\hbox{$\scriptscriptstyle\langle\!\langle$~}}
\def\fg{\leavevmode\raise.3ex\hbox{~$\!\scriptscriptstyle\,\rangle\!\rangle$}}

\usepackage{amsmath}
\usepackage{color}
\newcommand{\bea}{\begin{eqnarray}}
\newcommand{\eea}{\end{eqnarray}}
\newcommand{\be}{\begin{equation}}
\newcommand{\ee}{\end{equation}}

\DeclareMathOperator\atan{atan}
\DeclareMathOperator\ch{cosh}

\begin{document}

\begin{frontmatter}


\title{A liquid-gas transition for bosons with attractive interaction in one dimension}


\author[suny]{Christopher Herzog}
\author[umb]{Maxim Olshanii}
\author[lkb]{Yvan Castin}

\address[suny]{YITP, Department of Physics and Astronomy, Stony Brook University, Stony Brook, NY 11784, USA}
\address[umb]{Department of Physics, University of Massachusetts at Boston, Boston, MA 02125, USA}
\address[lkb]{Laboratoire Kastler Brossel, \'Ecole normale sup\'erieure, CNRS et UPMC, Paris, France}



\begin{abstract}
We consider a one dimensional system of $N$ bosons interacting via an attractive Dirac delta function potential.
We place the bosonic quantum particles at thermal equilibrium in a box of length $L$ with periodic boundary conditions.
At large $N$ and for $L$ much larger than the diameter of a two particle bound state, we predict by numerical and analytical studies of a 
simple model derived from first principles that the system exhibits a first order phase transition in a high temperature, non-degenerate regime.
The higher temperature phase is an almost pure atomic gas, with a small fraction of dimers, a smaller fraction of trimers, etc.  The lower temperature
phase is a mesoscopic or macroscopic bound state that collects all the particles of the system with the exception of a small gaseous fraction composed mainly of atoms.  
We term this phase, which is the quantum equivalent of the classical bright soliton, a liquid. 
\noindent 
\vskip 0.5\baselineskip

\end{abstract} 
\end{frontmatter}

\selectlanguage{english}
\section{Introduction and motivations}
\label{sec:intro}

In the last twenty years, ultracold atom experiments have been revealing their extreme flexibility in controlling
the parameters and the geometry. By freezing with a laser trap the atomic motion in its vibrational ground state 
along one or two spatial dimensions, one can prepare systems of effective reduced dimensionality, which has led
for example to the observation of the Berezinskii-Kosterlitz-Thouless transition \cite{JeanDalibard} in a two-dimensional Bose gas.
Furthermore, simply by applying a magnetic field close to a Feshbach resonance, one can adjust at will the $s$-wave scattering
length, that is the amplitude of the atomic interactions \cite{Feshbach}. The adiabatic use of both tools (transverse
confinement followed by a switch of the scattering length from positive to negative values) to an almost pure three-dimensional
Bose-Einstein condensate has allowed the experimentalists to observe the first $N$-mers in a one-dimensional atomic gas with attractive
interactions \cite{Salomon,Hulet}, that is of $N$-body bound states whose existence was predicted by the purely quantum theory 
for all $N>1$ \cite{McGuire}, and that correspond for large $N$ to the bright solitons of the classical field
(Gross-Pitaevskii or non-linear Schr\"odinger) equation.

In the present work, we identify another mechanism of formation of those $N$-mers, which is purely at thermal
equilibrium\footnote{The system is integrable, but it can be thermalized by contact with a buffer gas of a different atomic species.}
and in a totally different physical regime. We indeed show that a one-dimensional spatially homogeneous attractive Bose gas 
in a non-degenerate regime can be thermodynamically less favoured (that is of a higher free energy) than an almost pure
$N$-meric phase.
In other words, the non-degenerate gaseous phase may exhibit a first-order phase transition to an almost fully bound phase,
that we shall term a ``liquid" in a thermodynamic limit that we shall specify.

It turns out that the concept of a liquid-gas transition in a system of ultracold bosonic atoms is quite timely in three
dimensions, in the so-called unitary limit of an interaction with negligible range and infinite scattering length.
After the numerical discovery of bound states that are probably of a liquid nature \cite{vonStecher}, from $N=3$
(the Efimov trimers\cite{Efimov}) up to the largest accessible values of $N$ (a few tens),
such a transition was very recently observed in quantum Monte Carlo simulations with about one hundred particles \cite{Krauth}.
The experimental realisation, however, is hampered by strong three-body losses induced by the Efimov effect \cite{Salomon_pertes,Cornell},
and may require a dedicated experiment.

In this context, our study of the one-dimensional case has both a theoretical and a practical interest.
On one hand, the integrability of the theory allows one, under conditions to be specified in section \ref{sec:cdm},
to construct in a controlled way a model that is simple to study numerically and to interpret analytically,
as done in section \ref{sec:enei}, even in appropriate thermodynamic limits, see section \ref{sec:lim}.
On the other hand, the smallness of the three-body losses, here away from the unitary limit, opens the way to a short term
experimental realisation with ultracold atoms; we shall come back to this point in the conclusion.

\section{Construction of the model}
\label{sec:cdm}

We consider $N$ bosonic spinless quantum particles of mass $m$, living in one dimension on the $x$ axis in the absence of an external potential,
and with binary interactions via the attractive Dirac delta function potential $V(x_i-x_j)=g \delta(x_i-x_j)$, that is $g<0$.
A fundamental property of this hamiltonian model is its integrability, since one can determine the eigenstate wavefunctions
with the Bethe ansatz \cite{Gaudin}, the price to pay with respect to the usual repulsive case \cite{Lieb} being that one must
obviously include complex quasi-wavevectors \cite{Herzog,Caux} due to the existence
of negative energy eigenstates, as it was already pointed out in \cite{Lieb} and explicitly done for $N=2$ in an appendix of that reference.

\noindent{\bf In free space--} 
The Bethe ansatz solution is simple on the open line, that is in a pure scattering problem of the $N$ bosons; in this case,
the values of the quasi-wavevectors are known explicitly \cite{Herzog}. Under the condition that the $N$-body wavefunction
does not diverge at infinity, one finds that each eigenstate is composed of an arbitrary collection of indistinguishable
$n$-mers, the special case $n=1$ corresponding to an unbound bosonic particle, that is an atom.
Each $n$-mer is characterized by a well-defined total momentum $\hbar K$, and elastically scatters off the other $n$-mers, since
any chemical reaction (association or dissociation), or even any simple backscattering of two $n$-mers of different sizes,
is forbidden by integrability. For all $n\geq 2$, there exists one and only one possible bound state, of internal energy \cite{McGuire,Herzog}
\be
E_0(n) = -\frac{mg^2}{24 \hbar^2} n(n^2-1)
\label{eq:ener_liee}
\ee
In a given scattering state, there can be of course several $n$-mers with the same size, so that
the total number of fragments may vary from one (all the particles are bound in the ground $N$-mer,
as in a liquid) up to $N$ (all the particles exist under an atomic, gaseous form).
In terms of the size $n_i$ of the $i$th fragment and of its center of mass  wave vector $K_i$,
the scattering state energy may be simply written as the following sum of internal and kinetic energies:
\be
E = \sum_{i} \left[E_0(n_i) + \frac{\hbar^2 K_i^2}{2 n_i m}\right]
\label{eq:spectre}
\ee

\noindent{\bf In the box --} We shall here study the system at thermal equilibrium at temperature $T$, so we have to enclose
it in a quantisation box of length $L$, with the usual periodic boundary conditions\footnote{Even if it is possible to produce box potentials
\cite{Hadzibabic} it is useful to transpose our work to the harmonically trapped case, as was done in
the pioneering reference \cite{arxiv}.}.
The quasi-wavevectors of the Bethe ansatz then solve a non-linear system of coupled equations \cite{Caux} with no analytical
solution, which makes it difficult to determine the discrete energy spectrum in the box, even numerically, at large $N$.
The central idea of the present work is thus to restrict to a limiting case that may be handled in a simple way, such that
the free space expression (\ref{eq:spectre}) remains approximately true in the box, with (as the only twist) the natural
quantisation prescription:
\be
K_i \in \frac{2\pi}{L} \mathbb{Z}
\label{eq:quantif}
\ee

\noindent{\bf A box that is larger than the $n$-mers --} A first condition to get the expression (\ref{eq:spectre}) for the spectrum
is that the structure and the internal energy of the bound state are weakly affected by the presence of the box.
As the dimer state is the less strongly bound state, with the largest spatial diameter, this requires:
\be
\frac{L}{\ell} = \pi \left(\frac{|\mu_0|}{E_F}\right)^{1/2} \gg 1
\label{eq:cond1}
\ee
where $\ell\equiv2\hbar^2/(m|g|)$ is the diameter of the dimer, $\mu_0\equiv -m g^2 N^2/(8\hbar^2)$ is the chemical potential
of the $N$-mer, and $E_F=\hbar^2 (\pi \rho)^2/(2m)$ is the Fermi energy of the fictitious one-dimensional gas of fermions with the same density 
$\rho=N/L$ as the bosonic particles. As the $n$-mer wavefunction is an exponentially decreasing function
of the sum of the distances between the particles \cite{McGuire, Herzog}, we expect that neglecting the effect of the box
on the internal energy $E_0(n)$ introduces an exponentially small error in $L$. For $N=2$, this expectation is confirmed by the appendix A
of \cite{Lieb}: Two ``bound states" are found, corresponding to our dimer state with total momenta $\hbar K=0$ and
$\hbar K=2\pi\hbar/L$, and their energies deviate from (\ref{eq:spectre}) by $\simeq \pm 4 \exp(-L/\ell)$ in relative value.

\noindent{\bf In the non-degenerate regime --} The condition (\ref{eq:cond1}) guides all the present work.
Since $|\mu_0|$ is the binding energy of a particle in the $N$-mer, one can reasonably expect that a liquid-gas transition
(if there is any) takes place at a temperature of the order of a fraction of $|\mu_0|/k_B\gg E_F/k_B$, that is in a regime
where the atomic gaseous phase is highly non degenerate:
\be
\rho \lambda = \left(\frac{4 E_F}{\pi k_B T}\right)^{1/2} \ll 1
\label{eq:cond2}
\ee
where $\lambda=[2\pi \hbar^2/(m k_B T)]^{1/2}$ is the usual thermal de Broglie wavelength.
This atomic phase is then also close to the continuous spectrum limit, meaning that any discrete sum over the $K_i$
in equation (\ref{eq:quantif}) can be replaced by an integral, since $L \gg N \lambda$ implies $L\gg \lambda$.
In short, this phase can be considered as a classical gas.
In what follows we consider this high temperature limit, where the calculation of the external partition function $Z^{\rm ext}_{n\mathrm{\mbox{-}m\grave{e}res}}$ (that is of kinetic origin) of the $n$-mers associated to the spectrum (\ref{eq:spectre}) is greatly simplified. For a given
internal configuration of the $N$ particles, the subset of the $n$-mers forms a gas (supposed to be ideal for the moment)
with $\mathcal{N}_n $ indistinguishable (bosonic) elements of mass $n m\geq m$, de Broglie wavelength $\lambda/n^{1/2}\leq \lambda$
and density $\leq \rho$, so that this gas is as close to the classical limit as the ($n=1$) atomic gas, which leads to \cite{Diu}
\be
Z^{\rm ext}_{n\mathrm{\mbox{-}m\grave{e}res}}\simeq \left(\frac{Ln^{1/2}}{\lambda}\right)^{\mathcal{N}_n}\frac{1}{\mathcal{N}_n!}
\label{eq:Zext}
\ee

\noindent{\bf Negligible interactions among the $n$-mers --} 
A second validity condition for the use of the energy spectrum defined by Eqs.~(\ref{eq:spectre},\ref{eq:quantif})
is that the energy shifts in the box due to the {\sl elastic scattering}, i.e.\ interaction among $n$-mers, are negligible
as compared to the mean kinetic energy of each internal configuration\footnote{
Those elastic scattering events are for sure taken into account in the reference \cite{Herzog} in free space, but they
do not impact the spectrum (\ref{eq:spectre}) since the various $n$-mers are then asymptotically free.}. In the purely atomic phase,
the estimation of the interaction energy is made simpler by the following remark:
at the considered temperatures of order $|\mu_0|/k_B$, the kinetic energy per atom $\approx k_B T/2$ is much larger
than the dimer binding energy $|E_0(2)|$ as soon as $N\gg 1$,  which leads to
\be
\frac{\lambda}{\ell} = \frac{1}{N} \left(\frac{4\pi|\mu_0|}{k_B T}\right)^{1/2} \ll 1
\label{eq:cond3}
\ee
so that the typical relative wavevector of two atoms $\approx 1/\lambda$ is much {\sl larger} than $m|g|/\hbar^2$ ,
and two-body scattering takes place in the Born regime, see equation (99) in the reference \cite{Herzog}\footnote{
Contrarily to the three-dimensional case, the scattering of an incoming plane wave on a short range attractive potential in one dimension
always tends to its total reflection at low energy, which is beyond the Born regime.}.
The interaction energy per atom is then simply the one $\rho g$ of the mean field theory,
including the factor two due to the bosonic bunching effect, and it can be neglected if
\be
\frac{\rho |g|}{k_B T} = \frac{4}{\pi} \frac{(E_F |\mu_0|)^{1/2}}{N k_B T} 
= \frac{1}{\pi} (\rho \lambda) \left(\frac{\lambda}{\ell}\right)\ll 1
\label{eq:neg_int}
\ee
This conditions seems to be generically satisfied, similarly to (\ref{eq:cond3}), in some sort of thermodynamic limit, since $N$ 
appears in the denominator; this will be confirmed by section \ref{sec:lim}. Furthermore, this condition is implied
by the product of conditions (\ref{eq:cond2},\ref{eq:cond3}), so that it can be omitted in what follows.

To get an estimate of the interaction energy of an atom with a quasi-classical $n$-mer ($n\gg 1$), let us treat the corresponding bright
soliton with Bogoliubov theory, where the atom-$n$-mer scattering process at relative wave vector $k$ is described by the 
quasi-particle mode functions $u_k(x)$ and $v_k(x)$. If $u_k(x)\sim e^{i kx}$ for $k>0$ and $x\to-\infty$, the exact expression of the
modal functions \cite{Kaup,Sinha,soliton_interne} leads to $u_k(x)\sim e^{i \theta(k)} e^{ikx}$ for $x\to +\infty$, with a phase
shift $\theta(k)=-4\arctan (k\ell/n)$. The periodic boundary conditions then impose the quantization condition $\exp(i k L)\exp[i\theta(k)]=1$.
Hence in our big box $L\gg \ell/n$, a spacing between successive wave numbers equals $\delta k \simeq 2\pi/[L + \theta'(k)]$
where $\theta'(k)$ is the derivative of $\theta(k)$. The effective quantization length at a given thermal kinetic energy becomes
\be
L_{\rm eff} \approx L + \theta'(k=\sqrt{2\pi}/\lambda)
\label{eq:Leff}
\ee
We insert this effective length in the partition function  (\ref{eq:Zext}) written for a single atom, and we approximate
$\theta'(k)$ by its asymptotic expression $\sim -4n/(k^2\ell)$ at large $k$, which is justified by the fact that
$k_BT \approx |\mu_0|$. We then find a length change that is very small in relative value, and that corresponds, for an atom-$n$-mer
pair,  to an interaction energy of order $2 n |g|/L$, that is of order $n$ times the interaction energy of a pair of atoms\footnote{
It may be useful to refer to footnote \ref{note:test}.}.
Thus the same condition (\ref{eq:neg_int}) obtained
for interactions among atoms
allows us to neglect interactions between atoms and $n$-mers.
%

\noindent{\bf The model --} 
To conclude, for the validity conditions (\ref{eq:cond1},\ref{eq:cond2},\ref{eq:cond3}) given above, our high-temperature
system of $N$ bosonic particles with attractive interaction in a quantization box, may be assumed to exhibit, as internal configurations,
all the possible partitions of the $N$ particles into atoms and into two-body, \ldots, $N$-body bound states, each subset of atoms
or bound-states of a common size $n$ forming an ideal classical gas of $\mathcal{N}_n$ indistinguishable particles of mass $n m$,
which corresponds to the partition function (\ref{eq:Zext}). In mathematical terms, the full partition function
$Z$ of our model is the sum over all internal configurations with a fixed total number $N$ of particles, of 
configurational partition functions:
\be
Z = \sum_{\rm conf} Z_{\rm conf}, \ \ \mbox{where} \ \ \mathrm{conf}\in\{(\mathcal{N}_n)_{1\leq n\leq N}\in\mathbb{N}^N
\ \ \mbox{such that} \ \ \sum_{n=1}^{N} n\mathcal{N}_n = N\},
\label{eq:mod1}
\ee
that can be written, with $\beta=1/(k_B T)$, as products of internal and external partition functions,
\be
Z_{\rm conf}=\exp\left[-\beta \sum_{n=1}^{N} \mathcal{N}_n E_0(n)\right] 
\prod_{n=1}^{N} \left[\left(\frac{L n^{1/2}}{\lambda}\right)^{\mathcal{N}_n}\frac{1}{\mathcal{N}_n!}\right]
\label{eq:mod2}
\ee

\section{Mesoscopic numerical investigations and interpretation in terms of a liquid-gas transition}
\label{sec:enei}

\begin{figure}[htb]
\begin{center}
\includegraphics[width=5cm,clip=]{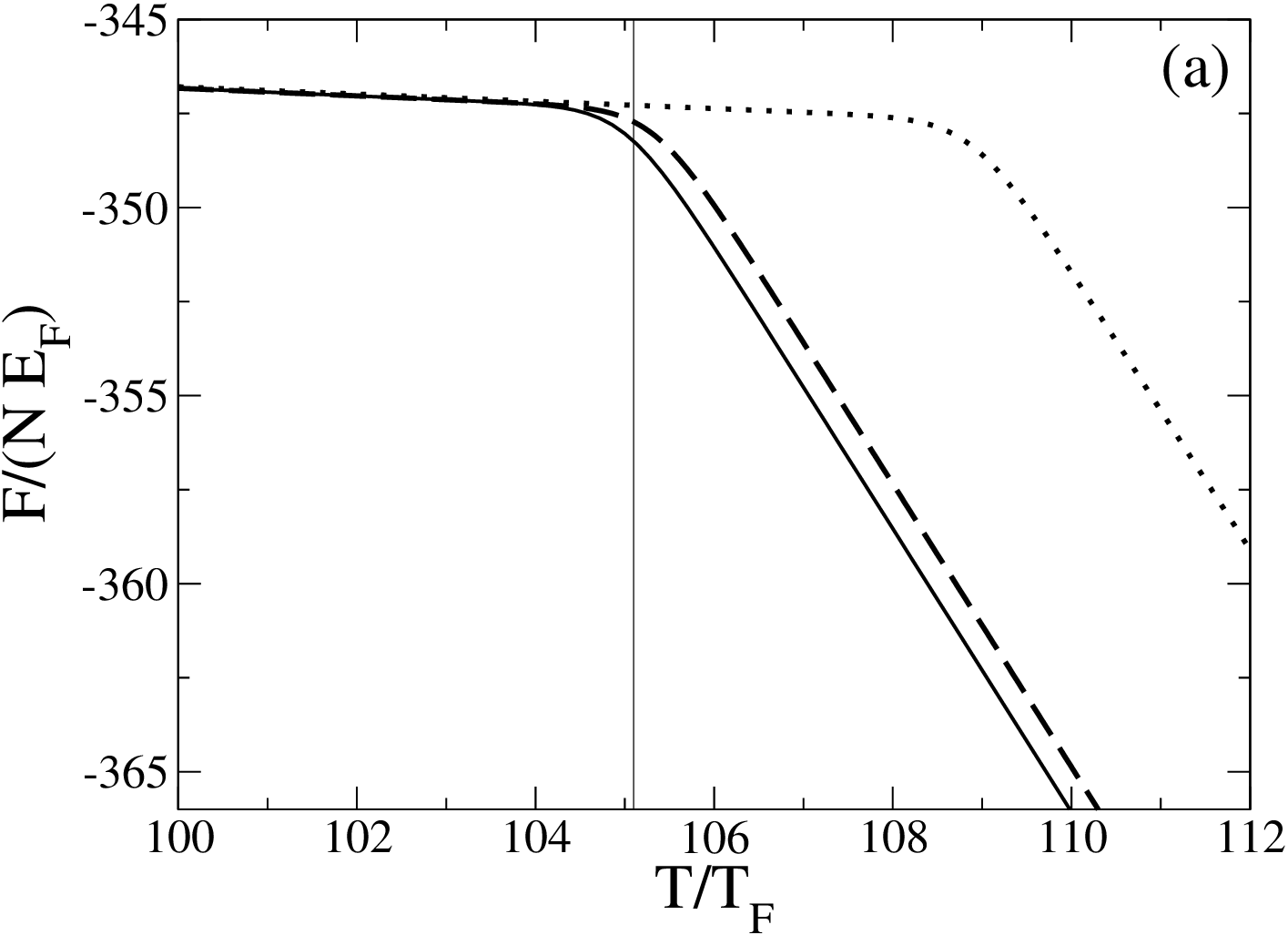} \hspace{2mm}
\includegraphics[width=5cm,clip=]{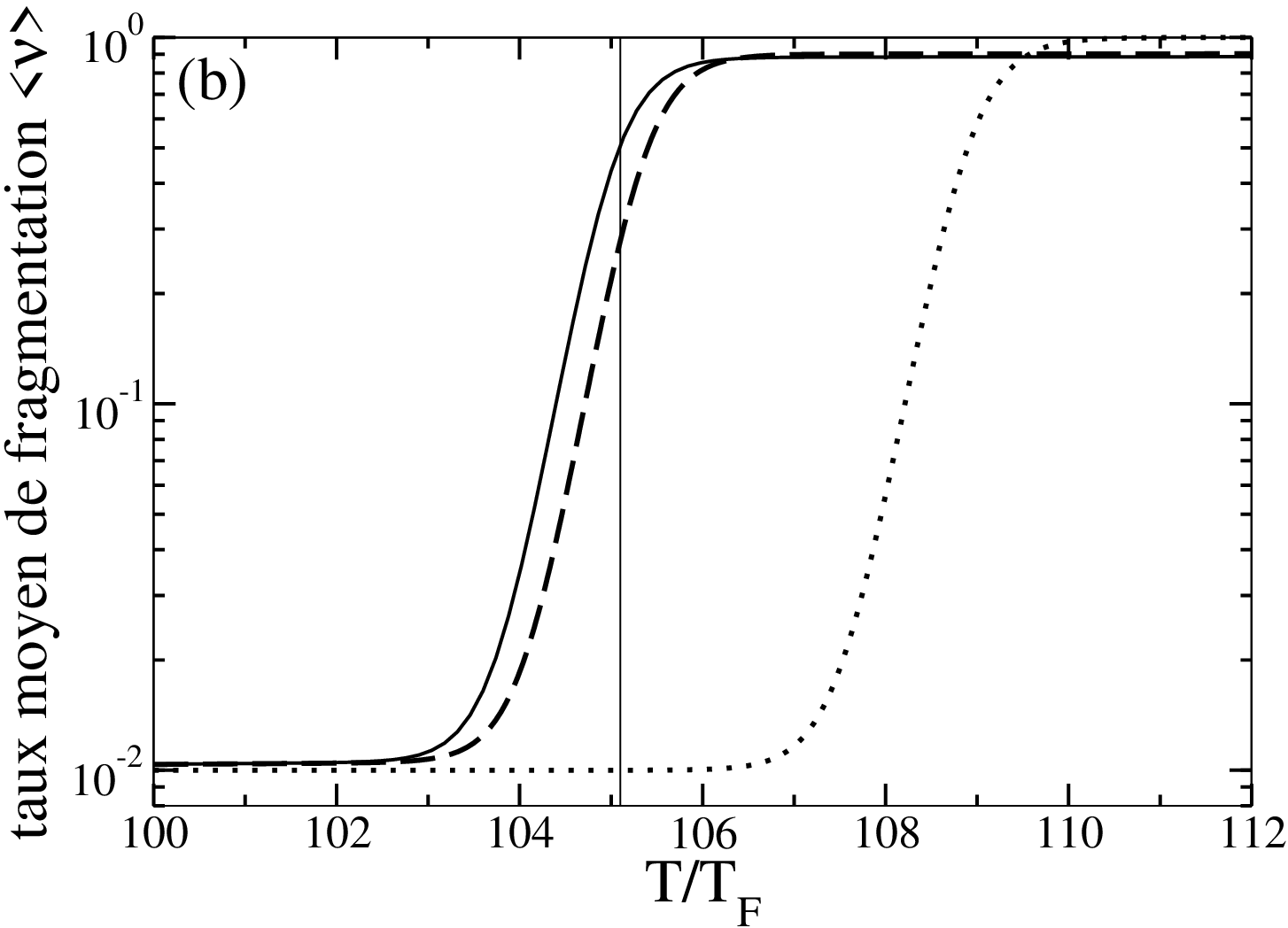} \hspace{2mm}
\includegraphics[width=5cm,clip=]{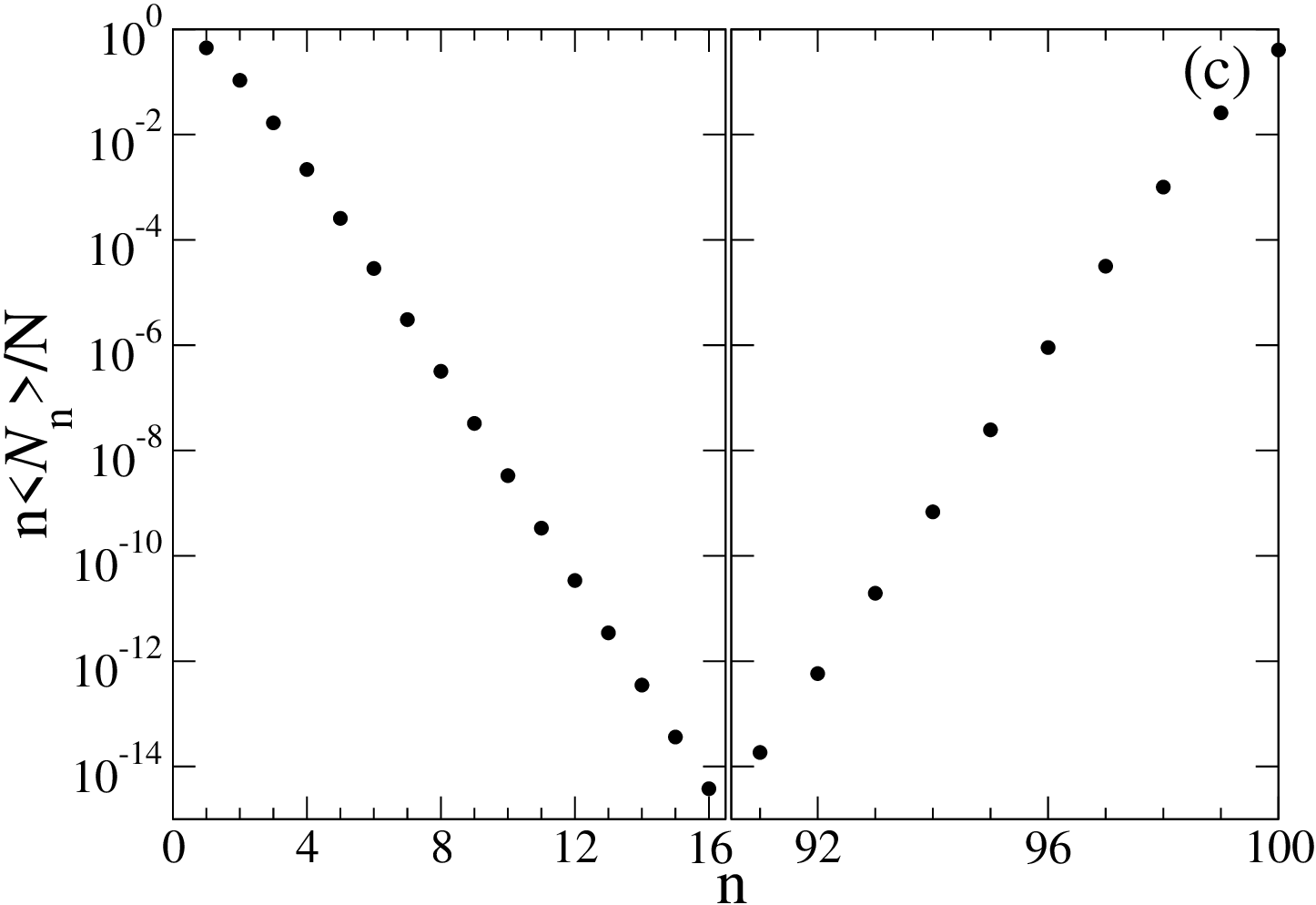}
\end{center}
\caption{For $N=100$ bosonic particles of mass $m$ in one dimension, with an attractive interaction potential $-|g| \delta(x_i-x_j)$
in a box of length $L=100 \ell$, where $\ell\equiv2\hbar^2/(m|g|)$ is the diameter of the two-body bound state:
(a) free energy per particle $F/N$, in units of the Fermi energy $E_F=k_B T_F=\hbar^2 (\pi \rho)^2/(2m)$,
and (b) mean fragmentation rate $\langle \nu\rangle$, as functions of the temperature in units of the Fermi temperature.
Solid line: the numerical result obtained by generation of all possible configurations.
Dotted line: when restricted to the two extreme configurations, the purely atomic one and the $N$-mer one.
Dashed line: further including the gaseous mixtures (in arbitrary proportions) of atoms and dimers, as well as the minimally
dissociated liquid phase of an atom plus a $(N-1)$-mer. In (c), the mean fraction of bosonic particles as a function of the size
of the $n$-mers, $n\mapsto n\langle \mathcal{N}_n\rangle/N$ (for all $n$ where it is $>10^{-15}$), numerically obtained at the
temperature $\approx 105.1 T_F$ where $\langle \nu\rangle$ crosses $(1+1/N)/2$ [vertical line in (a) and (b)]. Note
the logarithmic scale on the vertical axis in (b) and (c).
\label{fig:transi}
}
\end{figure}

We have performed a direct numerical study of the model (\ref{eq:mod1},\ref{eq:mod2}) simply by generating all possible
internal configurations $(\mathcal{N}_n)_{1\leq n\leq N}$. The numerical results are thus as exact as the model is.
The number of configurations to explore is, however, equal to the partition number $p(N)$  of the integer $N$, which
increases exponentially with $N$ \cite{Ramanujan}. This limits the simulations on a desktop computer
to $N\lesssim 150$. For a total number of $N=100$ particles and a fixed geometry $L/\ell=100$, which lies well within
the large box limit (\ref{eq:cond1}), we show in figure \ref{fig:transi}a and \ref{fig:transi}b the dependence on
temperature of two observables, the total free energy $F=-k_B T \ln Z$ and the mean fragmentation rate $\langle \nu\rangle$,
given by a thermal average of the relative number of fragments in each configuration with weights $Z_{\rm conf}$,
\be
\nu = \frac{1}{N} \sum_{n=1}^{N} \mathcal{N}_n \in \left[\frac{1}{N}, 1\right]
\ee
One observes in figure \ref{fig:transi}a a rather abrupt change of slope of the free energy, around $T\simeq 105 T_F\simeq 0.1 |\mu_0|/k_B$,
which is well within the non-degenerate regime (\ref{eq:cond2}). Since the derivative $-\partial_T F$ gives the entropy of
the system, this change of slope may be the precursor for our mesoscopic system of a first order phase transition with
a non-zero latent heat. Around the same temperature, there is a rather abrupt change of the mean fragmentation rate,
from a value close to unity at high temperature, to a value close to $1/N$, see figure
\ref{fig:transi}b: This change suggests the transition from a mainly atomic gaseous phase (i.e.\ an almost fully dissociated phase)
to a liquid phase mainly formed of the ground $N$-mer. This expectation is confirmed by the shape of the 
distribution $(n\langle \mathcal{N}_n\rangle/N)_{1\leq n\leq N}$ of the number of particles per size of the $n$-mers,
which exhibits at the transition two maxima, one at $n=1$ and the other at $n=N$, at the top of narrow peaks,
separated by infinitesimal occupation numbers for intermediate sizes, see figure \ref{fig:transi}c.
If one increases (or reduces) the temperature, one observes that the $n=N$ peak (the $n=1$ peak respectively) rapidly collapses
(not shown here).

Let us try to provide a more quantitative support to our liquid-gas transition scenario. At a first approximation,
let us keep in the partition function $Z$ only two internal configurations, the one with $N$ atoms, contributing
as $Z_{\rm at}$, and the one of the pure $N$-mer, contributing as $Z_{N\mbox{\scriptsize -m\`ere}}$. The corresponding
estimates for the free energy and the fragmentation rate, $F^{(0)}=-k_B T \ln(Z_{\rm at}+Z_{N\mbox{\scriptsize -m\`ere}})$
and $\langle \nu\rangle^{(0)}=(N^{-1} Z_{N\mbox{\scriptsize -m\`ere}}+Z_{\rm at})/(Z_{\rm at}+Z_{N\mbox{\scriptsize -m\`ere}})$,
are plotted as dotted lines in figure \ref{fig:transi}a and \ref{fig:transi}b. This reproduces well the general behavior of the numerical results,
up to about a 3.5\% shift towards higher temperatures, as we have checked explicitly.
The corresponding transition temperature $T_c^{(0)}$, obtained by equating the partition functions $Z_{\rm at}$ and 
$Z_{N\mbox{\scriptsize -m\`ere}}$, which amounts here to taking a mean fragmentation rate equal to $(1+1/N)/2$,
solves
\be
e^{-\beta_c^{(0)} E_0(N)} \frac{L\sqrt{N}}{\lambda_c^{(0)}} = \frac{1}{N!} \left(\frac{L}{\lambda_c^{(0)}}\right)^N
\label{eq:tc0}
\ee
For the parameters of figure \ref{fig:transi}, this leads to $T_c^{(0)}\simeq 108.90 T_F$, whereas the mean fragmentation rate
in the numerical simulation reaches the same value $(1+1/N)/2$ at $T\simeq 105.1 T_F$.

Note that a convenient approximate form of equation (\ref{eq:tc0}) is obtained by taking the logarithm
and then dividing by $N$. One thus takes the large $N$ limit, without precisely specifying yet how the thermodynamic
limit must be taken for $L$ (this will be specified in section \ref{sec:lim}). One only makes the weakly restrictive hypothesis
that $L/\lambda_c^{(0)}$ increases less rapidly than some power of $N$, so that $\ln(L/\lambda_c^{(0)})=O(\ln N)$. Using
Stirling asymptotic formula for $N!$, one gets
\be
\frac{1}{3}\beta_c^{(0)} |\mu_0| - \ln \left(\frac{e}{\rho \lambda_c^{(0)}} \right) = O\left(\frac{\ln N}{N}\right)
\label{eq:tc0impli}
\ee
Neglecting the terms that vanish at large $N$, and taking as unknown the dimensionless quantity $\beta_c^{(0)} |\mu_0|$,
one reaches the implicit equation
\be
\frac{2}{3}\beta_c^{(0)} |\mu_0| + \ln \left(\frac{2}{3} \beta_c^{(0)} |\mu_0|\right) = \ln \left(\frac{\pi e^2|\mu_0|}{6 E_F}\right)
\label{eq:Deltaimpli}
\ee
that can be solved in terms of the Lambert $W$ function (inverse function of $x\mapsto x e^x$ for $x\geq -1$) as\footnote{
To the same level of approximation, one predicts a latent heat per particle given by 
$Q^{(0)}/N=\frac{1}{3}|\mu_0|+\frac{1}{2}k_B T_c^{(0)}$,
since there is a transition from a $N$-mer of entropy 
$k_B O(\ln N)$ to an atomic classical gas of entropy $N k_B\ln[e^{3/2}/(\rho \lambda)]+k_B O(\ln N)$.}
\be
\frac{2}{3}\beta_c^{(0)} |\mu_0| = W\left(\frac{\pi e^2|\mu_0|}{6 E_F}\right)
\label{eq:tc0W}
\ee
According to the validity condition (\ref{eq:cond1}) of the model,
the argument of $W$ must be much larger than unity in equation (\ref{eq:tc0W}).
For the parameters of figure \ref{fig:transi}, this large $N$ approximation to $T_c^{(0)}$ reproduces the exact value
within only $3\%$ from below, even if the number of particles $N=100$ is not very large.

Let us now go beyond the two pure phases model and try to understand why its prediction (dotted line in figure \ref{fig:transi})
is slightly shifted along the temperature axis, with respect to the numerical results. At this stage, we can point
out another shortcoming of this model: at temperatures beyond the cross-over region, it assumes in practice that the system is purely
atomic with a unit fragmentation rate, whereas the numerical results present a persistent deviation of about
$10\%$ from full fragmentation. This last observation is consistent with a non-negligible population of dimers at the transition, see the
channel $n=2$ in figure \ref{fig:transi}c, and gives the idea of a second order approximation, that includes, in addition to the
pure atomic phase, arbitrary mixtures of atoms and dimers, so as to obtain an {\sl atomo-dimeric} phase\footnote{\label{note:S}
Its partition function is $Z_{\rm at-dim}=Z_{\rm at}S(z)$, where $z=\exp[-\beta E_0(2)] \lambda\sqrt{2}/L$ and the sum
over the number of dimers $s$, $S(z)=\sum_{s=0}^{[N/2]} z^s N!/[s!(N-2s)!]$, can be expressed in terms of an
hypergeometric function $S(z)={}_2 F_{0} (-\frac{N}{2},\frac{1}{2}-\frac{N}{2};4z)$.}.

To be symmetric, we also improve the competing liquid phase, even if this is quantitatively not really required
for the figure \ref{fig:transi}a and \ref{fig:transi}b (the dotted line is much closer to the solid line for $T\leq 105 T_F$
than for $T>105 T_F$): to the $N$-mer we add the {\sl minimally dissociated} configuration with one $(N-1)$-mer
and one atom, whose partition function is $\approx e^{-\beta |\mu_0|} \frac{L}{\lambda} Z_{N\mbox{\scriptsize -m\`ere}}$
for $N\gg 1$. Note here the occurrence of the expected thermal activation law exponential factor, which reveals
the fact that the $N$-mer state is protected from dissociation by an energy gap $|\mu_0|$
(in the center of mass frame) below the continuum of $N$-body excited states.
On the contrary, the minimally dissociated phase is favored by the kinetic entropic factor $\propto L/\lambda$
provided by the relative motion of the $(N-1)$-mer and the atom. In the particular case of figure \ref{fig:transi},
the effect of the energy gap largely dominates over the kinetic entropic effect, $e^{-\beta |\mu_0|} \frac{L}{\lambda} \approx 0.06$ 
at the transition, but this is not necessarily true for a large system where $L/\lambda$ can diverge.

For the extended phases that we have just described (the atomo-dimeric phase and the up-to-an-atom pure $N$-meric phase),
the free energy and the mean fragmentation rate are plotted as dashed lines on the figure \ref{fig:transi}a and
\ref{fig:transi}b. The dashed lines are significantly closer to the numerical results than the dotted line:
including the dimers has lowered the free energy of the gaseous phase and explains most of the shift in temperature.
The physics for this $N=100$ particle system thus seems to be quantitatively well understood.

We cannot be sure however that the phase extension performed here is sufficient in the limit of a large system.
In the gaseous phase, it remains to clarify analytically the role of dimers, or even trimers, quadrimers, etc; in the liquid phase,
as we have seen, inclusion of more severe dissociations of the $N$-mer may become necessary.
One can even imagine that a phase of totally different nature may emerge at large $N$. Can all this lead to large deviations
of the critical temperature from the simple estimate (\ref{eq:tc0}, \ref{eq:tc0W}), or even invalidate it? These questions
that cannot easily be answered by our mesoscopic numerical calculations are treated in the next section.

\section{Existence of the phase transition and various limits for a large number of particles}
\label{sec:lim}

According to the numerical simulations of the previous section, limited to a hundred of particles, our one-dimensional system of bosons with
attractive interaction should exhibit, in some macroscopic limit, a first order phase transition between a gaseous phase and a liquid phase,
with a critical temperature estimated within a two pure phases approximation, see equation (\ref{eq:tc0}). The goal of
this section is first to give an analytical proof of the existence of such a transition, with an upper bound on the error on
$T_c$ in (\ref{eq:tc0}); then to construct various types of thermodynamic limits, to see how they are influenced by finite size effects,
and to determine to what extent the two competing phases are mainly atomic or mainly liquid, respectively.

\subsection{Existence of the transition and bound on the error on $T_c$}
\label{subsec:prelim}
Appropriate bounds on the partition function $Z_{\rm conf}$ of equation (\ref{eq:mod2}) 
will allow one to show in an elementary way that deviations from monoatomicity in the gaseous phase and impurity
of the liquid phase do not destroy the scenario of a liquid-gas transition, provided that this transition takes place
in the non-degenerate regime $\rho \lambda\ll 1$. This excludes the possibility that a third phase, that would have been 
missed by the mesoscopic simulations, may emerge and dominate in the large $N$ limit. 
To this end, we shall define, in the configurational space, neighbourhoods of the purely atomic and $N$-meric phases,
and we shall bound the corrections that they bring to the free energy per particle, and hence to $T_c$, with respect to
the two pure phases approximation of the previous section.

\noindent{\bf In the neighbourhood of the monoatomic phase --}
Let us first take as a reference the purely atomic phase, with a partition function $Z_{\rm at}=(L/\lambda)^N/N!$,
by forming the ratio $Z_{\rm conf}/Z_{\rm at}$ and eliminating the atom number $\mathcal{N}_1$ thanks to the conservation
of the total number of particles $N$,
$\mathcal{N}_1 = N -\sum_{n=2}^{N} n \mathcal{N}_n$.
This elimination is made simpler by the expected property that $E_0(1)=0$ in equation (\ref{eq:ener_liee}). The only non trivial
remaining factor originates from the factorial due to atomic indistinguishability, that we shall simply bound,
\be
\label{eq:majutilfact}
\frac{N!}{\mathcal{N}_1!} = \left(\prod_{k=1+\mathcal{N}_1}^{N} k\right) \leq N^{N-\mathcal{N}_1}=N^{\sum_{n=2}^{N} n \mathcal{N}_n}
\ee
to obtain
\be
\frac{Z_{\rm conf}}{Z_{\rm at}} \leq \prod_{n=2}^{N} \frac{[\alpha_n(N)]^{\mathcal{N}_n}}{\mathcal{N}_n!}\ \ \ \mbox{with}\ \ \ 
\alpha_n(N)\equiv n^{1/2} e^{-\beta E_0(n)} N \left(\frac{N\lambda}{L}\right)^{n-1}
\label{eq:maj_zconf}
\ee
Intuitively, this inequality is almost saturated, and the integer numbers $\mathcal{N}_n$ for $n\geq 2$ almost obey a Poisson distribution
with parameter $\alpha_n(N)$ if a) the number of atoms in the considered phase only very weakly differs from $N$ in relative value
(in particular one should have $|1-\langle \mathcal{N}_1\rangle/N|\ll 1$), and b) the truncation effect due to conservation of the total number
of particles is negligible on the distribution of $\mathcal{N}_n$ (in particular one should have $\alpha_n(N) \ll N/n$).

A simple physical interpretation of the quantity $\alpha_n(N)$  is obtained as follows: The partition function of a gas
of $N-n$ atoms in the presence of a single $n$-mer, with $2\leq n\ll N$, is approximately equal to $\alpha_n(N) Z_{\rm at}$,
so that $\alpha_n$ is the first correction (in relative value) to the atomic gas partition function due to the $n$-mers.
In the expression for $\alpha_n(N)$, the factor $N^n \gg 1$ corresponds to an increase in the counting entropy, induced by the substitution of $n$ atoms by a
distinguishable $n$-mer. For a fixed $N$, the larger the size $n$ of the $n$-mer is, the smaller $\alpha_n(N)$ and the less
probable it is to include the $n$-mer, since $\rho\lambda \ll 1$; this already suggests that the trimers have a weaker effect on the free energy
of the gaseous phase than the dimers, and so on.

Let us now define the {\sl neighbourhood of maximal clustering order} $n_{\rm max}$ for the 
atomic phase, by the set of all possible configurations composed of $n$-mers of arbitrary sizes $n$ up to
$n_{\rm max}$. We shall take $n_{\rm max}$ less than $N/2$ so as to avoid overlap with the neighbourhood
of the $N$-mer to be defined soon. Our neighbourhood of maximal order $n_{\rm max}$ contains the purely atomic 
phase as a particular case, so that its partition function $Z_{\rm vois.\, at.}^{(n\leq n_{\rm max})}$ is bounded from
below by $Z_{\rm at}$. It can be bounded from above by applying (\ref{eq:maj_zconf}) and extending the summation over
the $(\mathcal{N}_n)_{2\leq n\leq n_{\rm max}}$ to $\mathbb{N}^{n_{\rm max}-1}$:
\be
Z_{\rm at} \leq Z_{\rm vois.\, at.}^{(n\leq n_{\rm max})}\equiv \!\!\!\! \sum_{{\rm conf}/\mathcal{N}_n=0\,\forall n> n_{\rm max}} 
\!\!\!\!\!\!\!\! Z_{\rm conf} \ \leq Z_{\rm at} 
\, e^{\sum_{n=2}^{n_{\rm max}} \alpha_n(N)}
\label{eq:zvoisat}
\ee
We shall use these inequalities to bound the change of critical temperature due to the presence of dimers, trimers, etc, 
in the gaseous phase. 

\noindent{\bf In the neighbourhood of the $N$-mer --}
Let us now take as a reference the pure $N$-mer, with a partition function $Z_{N\mbox{\scriptsize -m\`ere}}=\exp[-\beta E_0(N)]
N^{1/2}L/\lambda$. For $k< N-k$, that is $k<N/2$, let us define its {\sl subneighbourhood of dissociation degree} $k$ as the set of
configurations with one $(N-k)$-mer in presence of an arbitrary distribution of the remaining $k$ particles among atoms, dimers,
\ldots or at most one $k$-mer. Its partition function will be noted as $Z_{\mathrm{ss\mbox{-}vois.}\, N\mbox{\scriptsize -m\`ere}}^{(k)}$.
The particular case $k=0$ corresponds to the pure $N$-mer.

Let us then apply the previous upper bound reasoning, up to slight modifications: as one necessarily has $\mathcal{N}_{N-k}=1$,
and elsewhere $\mathcal{N}_n=0$ as soon as $n>k$, the number of atoms is now more conveniently eliminated by the relation
\be
\label{eq:sbq}
\mathcal{N}_1 = k-\sum_{n=2}^{k} n \mathcal{N}_n
\ \ \ \mbox{so that}\ \ \ 
\frac{k!}{\mathcal{N}_1!} = \left(\prod_{s=1+\mathcal{N}_1}^{k} s\right) \leq k^{k-\mathcal{N}_1}=k^{\sum_{n=2}^{k} n \mathcal{N}_n}
\ee
One also has $(N-k)^{1/2}\leq N^{1/2}$. All this leads to an upper bound for $Z_{\mathrm{ss\mbox{-}vois.}\, N\mbox{\scriptsize -m\`ere}}^{(k)}$
in the same spirit as (\ref{eq:maj_zconf}), except that the first corrections $\alpha_n$ (due to the $n$-mers) to the atomic phase surrounding the $(N-k)$-mer
now apply to $k$ particles only, rather than $N$, hence the substitution in equation (\ref{eq:maj_zconf}) of $\alpha_n(N)$ by $\alpha_n(k)$,
of $Z_{\rm at}$ by $(L/\lambda)^k/k!$ and of the upper bound $N$ by $k$:
\be
\frac{Z_{\mathrm{conf}}}{Z_{N\mbox{\scriptsize -m\`ere}}} \leq e^{-\beta[E_0(N-k)-E_0(N)]}
\frac{(L/\lambda)^k}{k!} \prod_{n=2}^{k} \frac{[\alpha_n(k)]^{\mathcal{N}_n}}{\mathcal{N}_n!}
\label{eq:ineg2}
\ee
Note that the first (exponential) factor is here a barrier to dissociation, according to the expected activation law, 
corresponding to the minimal energy $E_0(N-k)-E_0(N)>0$ required to extract $k$ particles from the $N$-mer, an energy
which is close to $k |\mu_0|$ for $k\ll N$. On the contrary, the powers of $L/\lambda\gg 1$ favor dissociation due to
a gain in kinetic entropy.
Another point is that, as in the discussion that follows equation (\ref{eq:maj_zconf}), the inequality (\ref{eq:ineg2})
is expected to be saturated when the product of dissociation is almost purely atomic, in which case the $n$-mers 
(with $n\geq 2$) are distributed according to Poisson laws of parameters $\alpha_n(k)$.
Let us now overbound the summation over all configurations $(\mathcal{N}_n)_{2\leq n\leq k}$ of the subneighbourhood of degree $k$
by extending this summation to $\mathbb{N}^{k-1}$, and let us define the $N$-mer {\sl neighbourhood of maximal degree of dissociation}
$k_{\rm max}$, by collecting all the subneighbourhoods of degree $k\leq k_{\rm max}$, which results in the partition function
\be
Z_{\mathrm{vois.}\, N\mbox{\scriptsize -m\`ere}}^{(k\leq k_{\rm max})} \equiv
\sum_{k=0}^{k_{\rm max}} Z_{\mathrm{ss\mbox{-}vois.}\, N\mbox{\scriptsize -m\`ere}}^{(k)}
\ee
We of course limit this definition to $k_{\rm max}<N/2$. We finally obtain the inequalities
\be
Z_{N\mbox{\scriptsize -m\`ere}} \leq Z_{\mathrm{vois.}\, N\mbox{\scriptsize -m\`ere}}^{(k\leq k_{\rm max})}
\equiv   \sum_{{\rm conf}/\exists n\geq N-k_{\rm max}/\mathcal{N}_{n}\neq 0}\!\!\!\!\!\! Z_{\rm conf}
\leq Z_{N\mbox{\scriptsize -m\`ere}} 
\sum_{k=0}^{k_{\rm max}} e^{-\beta[E_0(N-k)-E_0(N)]}
\frac{(L/\lambda)^k}{k!} e^{\sum_{n=2}^{k} \alpha_n(k)},
\label{eq:zvoiskmaxnm}
\ee
that we shall use to bound the critical temperature change due to partial dissociation of the $N$-mer.

\noindent{\bf Implications for the phase transition --}
Let us test the two pure phases scenario of section \ref{sec:enei}, by extending the first phase to the gaseous phase 
containing $n$-mers of arbitrary size up to $n_{\rm max}$, and the second phase to the liquid phase containing the partially
dissociated $N$-mer in the presence of an arbitrary phase with at most $k_{\rm max}$ particles.
The critical temperature $T_c^{(e)}$, where the exponent means ``extended",
is then obtained by requiring that the partition function $Z_{\rm vois.\, at.}^{(n\leq n_{\rm max})}$
of the extended gaseous phase, see equation (\ref{eq:zvoisat}), be equal to the partition function $Z_{\mathrm{vois.}\, N\mbox{\scriptsize -m\`ere}}^{(k\leq k_{\rm max})}$
of the generalized liquid phase, see equation (\ref{eq:zvoiskmaxnm}). The implicit equation (\ref{eq:tc0impli}) is replaced by
\be
\label{eq:tce}
\frac{1}{3}\beta_c^{(e)}|\mu_0| - \ln \left(\frac{e}{\rho \lambda_c^{(e)}} \right)
+O\left(\frac{\ln N}{N}\right)
=
\frac{1}{N} \left(\ln\frac{Z_{\rm vois.\, at.}^{(n\leq n_{\rm max})}}{Z_{\rm at}}
- \ln \frac{Z_{\mathrm{vois.}\, N\mbox{\scriptsize -m\`ere}}^{(k\leq k_{\rm max})}}{Z_{N\mbox{\scriptsize -m\`ere}}}\right)
\ee
In order to validate the two pure phases scenario, one has to show that the right-hand side of this equation
is small as compared to one of the first two terms of the left-hand side.
To this end, let us make the upper bounds more explicit in the sums over $n$ of the $\alpha_n(N)$ and the $\alpha_n(k)$,
using the fact that $-\beta E_0(n)/(n-1)=n(n+1) \beta |\mu_0|/(3 N^2)$, and that $n(n+1)$ is there at most $n_{\rm max}(n_{\rm max}+1)$
or $k_{\rm max}(k_{\rm max}+1)$:
\bea
\label{eq:majX}
\sum_{n=2}^{n_{\rm max}} \alpha_n(N) \leq N \sum_{n=2}^{n_{\rm max}} n^{1/2} X^{n-1} 
\leq N \frac{\sqrt{2} X}{(1-X)^2}\ \ \ &\mbox{with}&\ \ \ 
X=\frac{N\lambda}{L}\, e^{\beta |\mu_0| n_{\rm max} (n_{\rm max}+1)/(3 N^2)}\\
\label{eq:majY}
\sum_{n=2}^{k} \alpha_n(k)  \leq k \sum_{n=2}^{k_{\rm max}} n^{1/2} Y^{n-1}
\leq k \frac{\sqrt{2} Y}{(1-Y)^2}\ \ \ &\mbox{with}&\ \ \ 
Y=\frac{k_{\rm max} \lambda}{L}\, e^{\beta |\mu_0| k_{\rm max} (k_{\rm max}+1)/(3 N^2)}
\eea
For small enough values of $n_{\rm max}$ and $k_{\rm max}$, we expect that $X<1$ and $Y<1$, a property that we used 
to overbound by extending the summations up to infinity\footnote{We used $\sum_{n=2}^{+\infty} (n/2)^{1/2} u^{n-1}\leq
\sum_{n=2}^{+\infty} (n/2) u^{n-1} =u(1-u/2)/(1-u)^2\leq u/(1-u)^2$, $\forall u \in [0,1[$.}.
Insertion of (\ref{eq:majX}) in the inequalities (\ref{eq:zvoisat}) directly gives:
\be
0 \leq \frac{1}{N} \ln\frac{Z_{\rm vois.\ at.}^{(n\leq n_{\rm max})}}{Z_{\rm at}} \leq   \frac{\sqrt{2} X}{(1-X)^2}
\label{eq:majat}
\ee
To obtain a usable form of the upper bound (\ref{eq:zvoiskmaxnm}),
we note that $[E_0(N-k)-E_0(N)]/(k|\mu_0|)$ is a decreasing function of $k$ and is smaller than its value
$\eta=1-(k_{\rm max}/N)+(k_{\rm max}^2-1)/(3N^2)$ in $k=k_{\rm max}$. Then we use the upper bound in (\ref{eq:majY}).
Then we bound the resulting sum over $k$ by the corresponding series over $\mathbb{N}$:
\be
0 \leq \frac{1}{N} \ln \frac{Z_{\mathrm{vois.}\ N\mbox{\scriptsize -m\`ere}}^{(k\leq k_{\rm max})}}{Z_{N\mbox{\scriptsize -m\`ere}}} \leq 
\frac{e^{-\beta |\mu_0|\eta}}{\rho \lambda} e^{\sqrt{2}Y/(1-Y)^2} 
\label{eq:majliq}
\ee
To confirm the existence of a liquid-gas phase transition at a temperature close to $T_c^{(0)}$ defined in equation (\ref{eq:tc0}),
in the case $N\gg 1$, we maximally extend the two neighbourhoods, taking (for $N$ odd) $n_{\rm max}=k_{\rm max}=(N-1)/2$.
These maximal neighbourhoods are disjoint sets and they include as a whole {\sl all} the possible internal configurations\footnote{A configuration either
has no $n$-mer with a size larger than $N/2$, in which case it belongs to the maximal neighbourhood of the monoatomic gas;
or it has such a $n$-mer, and necessarily at most one, in which case the configuration belongs to the maximal neighbourhood of the $N$-mer.}.
Let us define $T_c$ as the temperature of equality of the partition functions of these maximal neighbourhoods, and let us
bound the supposedly small deviation of $T_c$ from $T_c^{(0)}$ by linearizing the left-hand side of equation (\ref{eq:tce}),
and by evaluating the right-hand side of that equation for $T=T_c^{(0)}$, which allows us to eliminate $\beta |\mu_0|$ in terms of $\rho\lambda$
in the quantities $X$ and $Y$ to obtain $X=2Y=e^{1/4} (\rho \lambda_c^{(0)})^{3/4}$.
We find that the upper bound in equation (\ref{eq:majat}) is larger than the one in (\ref{eq:majliq}).
As $\sqrt{2}e^{1/4}<2$, we conclude that there exists a neighbourhood of $\rho\lambda_c^{(0)}=0$,
that is of $L/\ell=+\infty$ according to (\ref{eq:tc0impli},\ref{eq:tc0W}), such that
\be
\frac{|T_c-T_c^{(0)}|}{T_c^{(0)}} < \frac{2 (\rho \lambda_c^{(0)})^{3/4}}{\ln[e/(\rho\lambda_c^{(0)})]}
\underset{\rho\lambda_c^{(0)}\to 0} \to 0
\label{eq:math}
\ee
Therefore there is indeed a liquid-gas transition in such a neighbourhood, in thermodynamic limits that we
now have to specify.
We note {\sl en passant} that the upper bound in (\ref{eq:majliq}), even if it is good enough in the present discussion
on the existence of the transition close to $T_c^{(0)}$, has the bad property of diverging as $1/(\rho \lambda)$
at very high temperature, so that it must be improved if one wishes to exclude the presence of a transition at temperatures arbitrarily higher than $T_c^{(0)}$. This is done in \ref{sec:mmvnmht}.

\subsection{A limit in which the model is exact}
\label{subsec:gentille}

As is apparent in the caption of figure \ref{fig:transi}, the figure only depends on two dimensionless parameters,
the number of particles $N$ and the box length scaled by the diameter of the dimer, $L/\ell$. 
To take the thermodynamic limit, the most natural way is to let $N$ tend as usual to infinity for fixed interaction strength
and mean density, which corresponds to
\be
N\to +\infty \ \ \mbox{with}\ \ \rho \ell = \mbox{constant}
\label{eq:lim1}
\ee
This makes the first validity condition (\ref{eq:cond1}) of the model exactly satisfied.

Let us first determine the critical temperature $T_c^{(0)}$ of the simplest, two pure phases approximation of section \ref{sec:enei}.
According to (\ref{eq:cond1}), the limit (\ref{eq:lim1}) implies that $(|\mu_0|/E_F)^{1/2}=N/(\pi\rho \ell) \to +\infty$;
this may look strange to the reader and we shall come back to this point in section \ref{subsec:mechante}.
By iteration of the implicit equation (\ref{eq:Deltaimpli}), taking into account the fact that $\beta_c^{(0)}|\mu_0|$ is much
larger than its logarithm, we obtain the asymptotic behavior:
\be
\frac{1}{3} \beta_c^{(0)}|\mu_0| \underset{N\to+\infty}{=} \ln N -\frac{1}{2} \ln\ln N + \ln \frac{e}{\rho\ell \sqrt{12\pi}} + O\left(\frac{\ln\ln N}{\ln N}\right)
\label{eq:devgen}
\ee
When $N$ diverges, the critical temperature $T_c^{(0)}$ thus very slowly becomes arbitrarily smaller than the chemical potential 
of the $N$-mer. Furthermore, the last two validity conditions (\ref{eq:cond2},
\ref{eq:cond3}) of the model become arbitrarily well satisfied at that temperature:
\be
\frac{\rho \lambda_c^{(0)}}{\rho\ell} = \frac{\lambda_c^{(0)}}{\ell}  
\underset{N\to+\infty}{\sim} \frac{(12\pi\ln N)^{1/2}}{N} \to 0
\ee
In other words, the simple model (\ref{eq:mod1},\ref{eq:mod2}) is asymptotically exact.
Last, due to the fact that  $\rho \lambda_c^{(0)}\to 0$ essentially as a power law, the equation (\ref{eq:math}) 
gives the important result that the two pure phases approximation is also asymptotically exact for the calculation
of $T_c$:
\be
(k_B T_c-k_B T_c^{(0)})/|\mu_0| =O(1/N^{3/4})\underset{N\to +\infty}{\to } 0
\ee
This does not imply that the dimers have a negligible contribution to the gaseous phase partition function, or that
their mean number in that phase tends to zero. From a study of the atomo-dimeric partition function, see section \ref{sec:enei},
or more simply by using the fact that the gaseous phase is here in the quasi-poissonian regime defined below equation (\ref{eq:maj_zconf}),
we find on the contrary that $Z_{\rm at\mbox{-}dim}/Z_{\rm at}$ diverges as $\exp[\alpha_2(N)]$, and that the mean number
of dimers diverges as $\alpha_2(N)$, where $\alpha_2(N)$ is defined in (\ref{eq:maj_zconf}) and diverges as $(\ln N)^{1/2}$ at the transition.
This is however $o(N)$ and too slow to asymptotically contribute to the free energy per particle or to the mean fragmentation rate.
Even better, the inclusion of trimers brings, according to (\ref{eq:maj_zconf}), a negligible contribution to the partition
function as compared to $Z_{\rm at}$, since this contribution is bounded by $\exp[\alpha_2(N)] \{\exp[\alpha_3(N)]-1\}\to 0$;
also, the mean number of trimers tends to zero as $\alpha_3(N)$ at the transition.
Similarly, the liquid phase is asymptotically pure there, because the probability that it contains an unbound atom tends to zero as $\exp(-\beta_c|\mu_0|)L/\lambda_c$,
that is as $\ln N/N$, see section \ref{sec:enei}.
As a consequence, one can determine the scaling law of the thermodynamic limit simply using the two pure phases approximation.
For example, one gets the following universal law for the mean fragmentation rate\footnote{%
One simply sets $T=T_c^{(0)}/(1-\epsilon)$, so that $Z_{\rm at}/Z_{N\mbox{\tiny -m\`ere}}=
\exp[-\ln(1-\epsilon)(N-1)/2]\exp[-\epsilon\beta_c^{(0)}E_0(N)]$, with $(N-1)\ln(1-\epsilon)\simeq -N\epsilon$ 
and $\beta_c^{(0)}E_0(N)\simeq -\beta_c^{(0)}|\mu_0|N/3$. Using (\ref{eq:devgen}) and a numerical opto-dimeric calculation
(cf.\ the dashed line in figure \ref{fig:transi}),
one finds for $\rho\ell=1$ that (\ref{eq:univ}) is already almost reached for $N=1000$.}~:
\be
\langle \nu\rangle \underset{N\to +\infty}{\to} \frac{\exp\delta}{1+\exp\delta} \ \ \ \mbox{\`a}
\ \ \ \delta\equiv \frac{T-T_c}{T_c} N \left(\frac{1}{2}+\frac{1}{3}\beta_c^{(0)}|\mu_0|\right)\ \ \mbox{fix\'e,}
\label{eq:univ}
\ee
This means that the relative temperature width of the cross-over region vanishes as $1/(N\ln N)$.

\subsection{A more usual thermodynamic limit}
\label{subsec:mechante}

The large $N$ limit of section \ref{subsec:gentille} is inappropriate for an experimental realisation at fixed
interaction strength, since $|\mu_0|$ then diverges as $N^2$ and the condition for the one-dimensionality of the system ($|\mu_0|$
small as compared to the transverse vibrational energy quantum $\hbar\omega_\perp$) is asymptotically unreachable at the transition
\footnote{This one-dimensionality condition also justifies the use of a Dirac $\delta$ function for the interaction potential, at least
in the usual cold atom case where the real three-dimensional interaction $V_{3D}$ has a negligible range as compared to the mean interparticle
distance \cite{Salomon}. It was indeed shown in \cite{Olshanii_g1d} that,
due to the transverse harmonic confinement, $V_{3D}$ induces in one-dimension an effective interaction of non-zero effective range $r_e\approx
[\hbar/(m\omega_\perp)]^{1/2}$. This effective range is negligible in the Born regime under the sufficient condition $k_{\rm rel}|r_e|\ll 1$
where the relative wavevector is here $k_{\rm rel}=O(N/\ell)$.}.
This gives the idea of fixing the value of $N g$ to a constant in the experiment, by tuning $g$ with a Feshbach resonance.
The limit (\ref{eq:lim1})  then corresponds to $N^2/L=$constant, which is unusual. Let us rather consider here the limit
\be
N\to +\infty \ \ \ \mbox{with}\ \ \ \frac{L}{\ell}=\mbox{constant}
\label{eq:lim2}
\ee
that indeed leads to a constant density $\rho$ at fixed $N g$, and to a constant ratio $|\mu_0|/E_F$ as expected
in a regular thermodynamic limit. Then $\beta_c^{(0)}|\mu_0|$ has a finite limit, given by (\ref{eq:tc0W}),
and $\rho\lambda_c^{(0)}/e=\exp(-\beta_c^{(0)}|\mu_0|/3)$.
However, $T_c^{(0)}$ no longer coincides with $T_c$; it is only a good approximation in a sufficiently non degenerate
regime, as guaranteed by the mathematical result (\ref{eq:math})\footnote{
The more precise forms (\ref{eq:majat},\ref{eq:majliq}) guarantee the existence of a solution $T_c$ to the equation
$Z_{\mathrm{vois.}\, N\mbox{\tiny -m\`ere}}^{(k\leq N/2)}(T)=
Z_{\rm vois.\, at.}^{(n\leq N/2)}(T)$ as soon as $L/\ell>105.1$.}.
The same conclusion holds for the universal relation (\ref{eq:univ}), that now predicts a width of the cross-over
region scaling as $1/N$.
The first two validity conditions (\ref{eq:cond1},\ref{eq:cond2}) of the model are now only approximately 
obeyed; only the third condition (\ref{eq:cond3}) is exactly fulfilled when $N\to +\infty$ at the transition.

Even if the mathematical result (\ref{eq:math}) has to consider the maximal neighbourhoods of the two pure
phases, with $n_{\rm max}\sim k_{\rm max}\sim N/2$, to be rigorous, it is physically expected that the first corrections
to $T_c^{(0)}$ actually originate from narrower neighbourhoods $n_{\rm max}\ll N/2$ et $k_{\rm max}\ll N/2$.
As the close neighbourhood of the purely atomic gas is quasi-poissonian at the transition, as defined below equation (\ref{eq:maj_zconf}),
we find that its leading contribution to the right-hand side of (\ref{eq:tce}) is the one  $\alpha_2(N)/N\to \sqrt{2} \rho\lambda$ of 
the dimers\footnote{To go beyond the poissonian approximation, we have applied the method of Laplace to the sum $S(z)$
of the footnote \ref{note:S}, after having used Stirling's formula and replaced the sum over $s$ by an integral over
$x=2s/N$. Then $Z_{\rm at-dim}/Z_{\rm at}\sim \exp[N u(x_0)]/(1+x_0)^{1/2}$ when $N\to +\infty$, with
$u(x)=(x/2)\ln[2\sqrt{2}\rho\lambda/(e x)]-(1-x)\ln(1-x)$ and $x_0$ is the root of $u'(x)$ in the interval $[0,1]$.
One always has $u(x_0)<\sqrt{2}\rho \lambda$, in agreement with the upper bound (\ref{eq:zvoisat}) applied to 
$n_{\rm max}=2$, and $u(x_0)\sim \alpha_2(N)/N$ if $\rho\lambda\to 0$.}, the contribution of the trimers $\alpha_3(N)/N$
being $O(\rho\lambda)^2$, etc. We realize however that there is a {\sl limitation} to our model: although it is non
degenerate, the purely atomic phase remains in reality a bosonic gas, and its partition function $Z_{\rm at}^{\rm Bose}$
already deviates from the classical gas one $Z_{\rm at}=(L/\lambda)^N/N!$ to first order in $\rho \lambda$ \cite{Diu}, $N^{-1} \ln (Z_{\rm at}^{\rm Bose}/Z_{\rm at})=2^{-3/2} \rho \lambda
+O(\rho\lambda)^2$, which brings a correction to the critical temperature to the same order in $\rho\lambda$ as
the one of the dimers, but which is ignored by the model.

To estimate the contribution of the close neighbourhood of the $N$-mer, we take inspiration from the above discussion 
and assume, to leading order in $\rho\lambda$, that the $N$-mer is weakly dissociated in a purely atomic gas with $k$ particles, well in the poissonian
regime. This amounts to saying that the second inequality in (\ref{eq:zvoiskmaxnm}) is almost saturated, 
that the $\alpha_n(k)$ are negligible, and that the $N$-mer dissociates into a $(N-k)$-mer plus $k$ particles with an essentially
poissonian weight, since one takes $k_{\rm max}\ll N/2$:
\be
Z_{\mathrm{ss\mbox{-}vois.}\, N\mbox{\scriptsize -m\`ere}}^{(k)} /Z_{N\mbox{\scriptsize -m\`ere}} \simeq \frac{\gamma^k}{k!}
\ \ \ \mbox{where}\ \ \ \gamma = N \frac{e^{-\beta |\mu_0|}}{\rho \lambda}
\label{eq:poissliq}
\ee
At the transition, the mean dissociation rate of the liquid phase is then $\gamma/N\simeq (\rho \lambda_c^{(0)})^2/e^3$,
which brings to the right-hand side of (\ref{eq:tce}) a contribution negligible as compared to the one of the dimers of the
gaseous phase.

\noindent{\bf Beyond the model --} In order to determine the true correction to $T_c^{(0)}$ to first order in $\rho\lambda$,
let us apply to the gaseous phase\footnote{
\label{note:test} In the liquid phase, one can apply Bogoliubov theory to the bright soliton if the Bogoliubov
quasi-particles, of spectrum $\epsilon_k = |\mu_0|+ \hbar^2 k^2/(2m)$, are a number $\delta N =\sum_k [\exp(\beta \epsilon_k)-1]^{-1} \ll N$,
which imposes $k_B T \ll |\mu_0|$  if $\rho \lambda\ll 1$, that is $\delta N/N \simeq e^{-\beta |\mu_0|}/(\rho\lambda)\ll 1$:
in the limit (\ref{eq:lim2}) for fixed $Ng$, the corresponding partition function 
obeys $N^{-1}\ln Z_{\rm Bog}=\frac{1}{3} \beta |\mu_0| + (\rho \lambda)^{-1} \sum_{s\geq 1}\exp(-\beta |\mu_0| s)/\sqrt{s}$,
with a leading $s=1$ term, for $k_B T \ll |\mu_0|$, that is indeed correctly given by (\ref{eq:poissliq}); 
the contribution to $N^{-1} \ln Z_{\rm Bog}$ of the phase shift $\theta(k)=-4\atan(k\ell/N)$ due to the soliton [see above (\ref{eq:Leff})]
indeed tends to zero at large $N$, as $-(4\ell/L)\exp(-\beta |\mu_0|)[1+O(k_B T/|\mu_0|)]/(N\rho \lambda)$.}
a systematic approach at high temperature, that is the quantum virial expansion,
$N^{-1} \ln (Z_{\rm gaz}^{\rm Bose}/Z_{\rm at})=b_2 \rho \lambda+ O(\rho\lambda)^2$,
that includes at this order quantum statistics, atomic interactions and the presence of dimers \cite{Landau}.
With the method of \S 2.3 of reference \cite{canadien}\footnote{
As in \cite{Ouvry,Drummond} we use a harmonic regulator of angular frequency $\omega$, so that
$b_2/\sqrt{2}=1/4+\lim_{\omega\to 0} \sum_{q\geq 0} (e^{-\beta \epsilon_q}
-e^{-\beta\epsilon_q^{(0)}})$ where $(\epsilon_q)_{q\in\mathbb{N}}$ is the spectrum of $H=p^2/m +m\omega^2 x^2/4+g\delta(x)$
restricted to the even (bosonic) states, solving $\Gamma(\frac{1}{4}-\frac{\epsilon_q}{2\hbar\omega})/\Gamma(\frac{3}{4}-\frac{\epsilon_q}{2\hbar\omega})
=-2(2\hbar\omega/m)^{1/2}\hbar/g$ \cite{Wilkens} and $\epsilon_q^{(0)}=(2q+1/2)\hbar\omega$ is the corresponding $g=0$ spectrum.
If $\omega\to 0$ for a fixed $g<0$, $\epsilon_0\to E_0(2)$ and $\epsilon_{q\geq 1}=\hbar\omega [2q-1/2+2\Delta(\epsilon_q)+O(1/q)]\geq 0$,
with $\Delta(\epsilon)\equiv (1/\pi) \arctan\{[-\epsilon/E_0(2)]^{1/2}\}$. If $g>0$, $\epsilon_{q\geq 0}=\hbar\omega[2q+3/2-2\Delta(\epsilon_q)+O(1/q)]\geq 0$
with the same function $\Delta(\epsilon)$. It remains to replace the sum over $q$ by an integral over $\epsilon_q$, to obtain $b_2$
as given in the text for $g<0$, and $b_2/\sqrt{2}=\frac{e^u}{2}[1-\mbox{erf}\,(u^{1/2})]-\frac{1}{4}$ (in agreement with \cite{jmath})
for $g>0$.}
we find for attractive interactions $b_2/\sqrt{2}=(e^u/2)[1+\mbox{erf}\,(u^{1/2})]
-1/4$, where $u\equiv \beta |E_0(2)|$; in the limit (\ref{eq:lim2}), $u\to 0$ so that $b_2\to 2^{-3/2}$,
which reduces to the effect of quantum statistics, contrarily to the predictions of the model,
and makes the temperature shift attributed to the dimers in figure \ref{fig:transi} not meaningful\footnote{
On the contrary, the fraction of dimers predicted by the model in the gaseous phase agrees with the one $\sqrt{2}\rho\lambda \exp[-\beta E_0(2)]$
deduced from $b_2$.}!
We interpret this as being due to the orthogonality of the diatomic unbound states with the dimer state, a fact not taken into account
in the model; if the coupling constant $g$ was changing from $0^+$ to $0^-$, the dimer state would appear but a diatomic unbound state
would disappear, at the energy scale $|E_0(2)| \ll k_B T$, leading to a compensation in $b_2$: the second cluster coefficient
$b_2$ indeed varies continuously around $g=0$. In our {\sl corrected} model,
we finally obtain the following implicit equation for the critical temperature of the liquid-gas transition 
in the limit (\ref{eq:lim2}):
\be
\frac{1}{3}\beta_c |\mu_0| \underset{\mbox{\small model}}{\stackrel{\mbox{\small corrected}}{=}} \ln \left(\frac{e}{\rho \lambda_c} \right) 
+ 2^{-3/2} \rho \lambda_c +O(\rho\lambda_c)^2
\label{eq:tccorrigee}
\ee

\section{Conclusion}
\label{sec:conclusion}

We have here restricted our study of the unidimensional system of attractive ($g<0$) bosons to the particular case of a big box, with a length
$L$ much larger that the diameter $\ell$ of the two body bound state. This means that, in
a uniform distribution of the $N$ bosons at the density $\rho=N/L$ in the box,
the mean interparticle spacing $1/\rho$ is much larger that the ground $N$-mer of
diameter $\xi\simeq\hbar^2/(N m|g|)$ and density profile $[N/(4\xi)]/\ch^2[x/(2\xi)]$ at large $N$.
In short, we are in the regime $\rho \xi = \ell/(2L) \ll 1$.

In this particular regime and at thermal equilibrium, we have shown that
a liquid-gas transition occurs in the thermodynamic limit, at a temperature $T_c$ given to various orders of approximation by 
(\ref{eq:tc0W},\ref{eq:tccorrigee}), and so high that the system lies in the non-degenerate regime
$\rho\lambda_c \ll 1$, where $\lambda_c$ is the corresponding thermal de Broglie wavelength.
Over an interval of temperature of width $O(1/N) T_c$, the system changes from an essentially monoatomic uniform gaseous phase of density
$\rho$ [with a $O(\rho \lambda_c)^n$ fraction of $n$-mers], to a
liquid phase in the form of a single inhomogeneous drop of diameter $\xi$, surrounded by a small $O(\rho\lambda_c)^2$ 
gaseous fraction which is mainly monoatomic and uniform.

On a theoretical point of view, our result seems to contradict a general argument due to Landau, 
that precludes the coexistence of two different phases
in unidimensional systems with short range interactions, due to the fact that
these phases have a tendency to experience arbitrarily high fragmentation
and to form an emulsion \cite{Landau}. However the $n$-mers here do not constitute an
ordinary extensive liquid with a density bounded from above: their energy $E_0(n)$ is
not a linear function of the number of particles $n$ at large $n$; it is rather a cubic function, 
so that the splitting of the liquid in mesoscopic or macroscopic
fragments has a very high energy cost (for example $(3/4)|E_0(N)|$ 
if one cuts the $N$-mer into two halves), and it is not simply proportional to
the number of fragments with an intensive proportionality factor as it was
supposed in \S 163 of reference \cite{Landau}.

At this stage, one may formulate the following objection: The lack of extensivity of the predicted liquid phase results from the Dirac
delta function interaction potential, which should become unrealistic when $N\to +\infty$ since the liquid density then (presumably)
diverges. In this limit, any realistic model for the interaction is expected to exhibit some finite range effect, such as a hard-core contribution.
In the thermodynamic limit, however, it is the dimensionless quantity $L/\ell$ that matters, see equation (\ref{eq:lim2}), rather
than the length $\ell$. The objection can thus be circumvented by adjusting the interaction strength with the particle number $N$,
thanks to a Feshbach resonance for ultracold atoms, in such a way that the central liquid density $N/(4\xi)$ remains constant.
This requires that $N^2/\ell$ remains constant, so that the coupling constant $g$ has to vanish as $1/N^2$.
According to (\ref{eq:lim2}), the box length $L$ then diverges as $N^2$, and the ground liquid phase chemical potential $\mu_0$,
the gaseous phase Fermi energy $E_F$ and the transition temperature $T_c$ all vanish as $1/N^2$.
At the transition temperature, however, the quantum degeneracy parameter $\rho \lambda_c$ remains fixed, and 
so does the entropy per particle in the gaseous phase, which shows that these low $\propto 1/N^2$ temperatures can in principle be 
reached by adiabatic cooling.

The lack of extensivity of the liquid phase may have an important consequence
on the metastability of the gaseous phase below the transition temperature.
In general, it is known that a slowly cooled macroscopic gas in principle
presents a very large delay to liquefaction, due to the extensive free energy
barrier between the two phases, but in practice, this is avoided by local
nucleation of droplets. Here this local mechanism may be inhibited by the
strong non linearity of $E_0(n)$ with $n$. Let us indeed isolate in thought a small
fraction of the gaseous phase with $N'$ particles, with $1,\rho \ell \ll N'\ll N$,
which
is allowed by its homogeneity and the absence of long range interaction.
As the liquefaction temperature $T_c$ of equation (\ref{eq:tc0impli}) is, within logarithmic
corrections, proportional to the binding energy of a particle in the liquid
that forms, it becomes obvious that the small fraction shall liquefy at a
temperature $T_c'\propto N'^{2}$ much smaller than the one $\propto N^2$ of the whole gas.

If it happens that this metastability constitutes a serious experimental
handicap, a way out could be to measure the free energies of the two phases
on both sides of the critical temperature, using a gradual heating of the
already observed liquid phase at low temperature \cite{Salomon,Hulet} and a slow cooling of
the easy-to-prepare gaseous phase at high temperature, and then to realise
that the free energies cross at $T_c$. The free energy $F(T)$ can indeed be deduced
from the mean energy $E(T)$ by integration of the relation $E=\partial_{\beta}(\beta F)$, where $\beta=1/(k_B T)$. 
The mean energy, as it was shown in reference \cite{relgen}, is a known
functional of the momentum distribution $n(k)$ of the particles, that can be
measured by time of flight after abrupt switch-off of the interactions thanks
to a Feshbach resonance.

All this allows us to reasonably hope that an experimental test of our scenario is possible, 
maybe leading to the observation of a liquid-gas transition
with ultracold atoms, which would constitute a ``grande premi\`ere".



\section*{Acknowledgments}
The group of Y.C. is also affiliated to IFRAF. One of us (C.H.) has received a Fulbright grant
for his 1997-1998 stay in Paris, where he has accomplished most of this work, as well as
National Science Foundation grants PHY-1316617 and 0844827; he also thanks the Sloan Foundation. 
M.O.\ was supported by grants from the Office of Naval Research
(N0004-121-0400) and the National Science Foundation (PHY-1019197).
We thank Christoph Weiss for his incitation to finalize this project.

\appendix
\section{An improved upper bound for the $N$-mer neighbourhood at very high temperature}
\label{sec:mmvnmht}

In the main text, see equation (\ref{eq:majliq}), we have obtained the following upper bound of the partition function
of the $N$-mer neighbourhood of maximal degree of dissociation $k_{\rm max}<N/2$:
\begin{equation}
\frac{1}{N} \ln Z_{\mathrm{vois.} N\mbox{\scriptsize -m\`ere}}^{(k\leq k_{\rm max})}
\leq \frac{1}{N} \ln Z_{N\mbox{\scriptsize -m\`ere}} + \theta \ \ \mbox{with}\ \ \theta=\frac{e^{-\beta |\mu_0| \eta}}{\rho
\lambda} e^{\sqrt{2}Y/(1-Y)^2}
\label{eq:maj_anc}
\end{equation}
This upper bound is useful close to the critical temperature but it is not in the limit of very high temperatures,
in particular $T\gg T_c^{(0)}$. When $\rho \lambda\to 0$ indeed, $\beta |\mu_0|$ tends to zero, as well as
$Y$, according to the equality
\begin{equation}
\beta |\mu_0| = \frac{1}{4\pi} (\rho\lambda)^2 (L/\ell)^2
\end{equation}
Then $N^{-1} \ln Z_{N\mbox{\scriptsize -m\`ere}}$ tends to zero as $(\rho \lambda)^2$, but $\theta$ diverges as $1/(\rho \lambda)$, that is
faster  that $N^{-1} \ln Z_{\rm at}$, which grows logarithmically only, as $\ln[e/(\rho \lambda)]$.
The upper bound (\ref{eq:maj_anc}) is thus not sufficient to mathematically exclude the existence of a transition other
than the one discussed in this paper, at a higher temperature.

The existence of such a competing transition can fortunately be excluded by a simple improvement of the upper
bound in (\ref{eq:maj_anc}). One simply has to stop before the last step having led to (\ref{eq:majliq}), that is one keeps
the finite sum over $k$:
\begin{equation}
\label{eq:avant-der}
\frac{1}{N} \ln \frac{Z_{\mathrm{vois.} N\mbox{\scriptsize -m\`ere}}^{(k\leq k_{\rm max})}}
{Z_{N\mbox{\scriptsize -m\`ere}}} \leq \frac{1}{N} \ln \sum_{k=0}^{k_{\rm max}} \frac{(N \theta)^k}{k!}
= \frac{1}{N} \ln \left[\frac{e^{N\theta}}{k_{\rm max}!} \Gamma(1+k_{\rm max},N\theta)\right]
\end{equation}
where $\Gamma$ is the incomplete Gamma function. The integral representation of this function leads to the following discussion,
for $N\to +\infty$ at fixed $\theta$ and $k_{\rm max}/N$, which corresponds to the usual thermodynamic limit of section \ref{subsec:mechante}.
If $N \theta/k_{\rm max}<1$, an equivalent of the upper bound in (\ref{eq:avant-der}) is indeed obtained by extending the sum over
$k$ to the entire $\mathbb{N}$, and one recovers (\ref{eq:majliq}). If $N \theta/k_{\rm max}>1$, an equivalent is obtained by the upper bound
\begin{equation}
\sum_{k=0}^{k_{\rm max}} \frac{(N \theta)^k}{k!} = \frac{(N \theta)^{k_{\rm max}}}{k_{\rm max}!}
\sum_{k'=0}^{k_{\rm max}} \frac{1}{(N \theta)^{k'}} \frac{k_{\rm max}!}{(k_{\rm max}-k')!}
\leq \frac{(N \theta)^{k_{\rm max}}}{k_{\rm max}!}
\sum_{k'=0}^{+\infty} \left(\frac{k_{\rm max}}{N\theta}\right)^{k'}
\end{equation}
where we have set $k'=k_{\rm max}-k$, we have used an upper bound of the type (\ref{eq:majutilfact})  and we have extended
the summation over $k'$ to $\mathbb{N}$. The logarithm of the resulting geometric series is bounded and has a vanishing contribution
in the thermodynamic limit. In other words, depending on the fact that $N \theta/k_{\rm max}$ is less than or larger than one,
the sum over $k$ in (\ref{eq:avant-der}) is dominated by its last or its first terms, at large $N$.
We finally keep:
\begin{equation}
\label{eq:maj_nouv}
0 \leq \frac{1}{N} \ln \frac{Z_{\mathrm{vois.} N\mbox{\scriptsize -m\`ere}}^{(k\leq k_{\rm max})}}{Z_{N\mbox{\scriptsize -m\`ere}}}
\leq \left\{
\begin{tabular}{ccc} $\theta$ & \mbox{if} & $N\theta/k_{\rm max} <1$ \\
$\frac{k_{\rm max}}{N} \ln (eN \theta /k_{\rm max})$ & \mbox{if} & $N\theta/k_{\rm max} > 1$
\end{tabular}
\right.
\end{equation}
In the limit $\rho\lambda\to 0$, the upper bound in  (\ref{eq:maj_nouv}) diverges
logarithmically only, with a prefactor $k_{\rm max}/N <1/2$, that is at least two times less rapidly than $N^{-1}\ln Z_{\rm at}$.

\begin{figure}[t]
\centerline{\includegraphics[width=10cm,clip=]{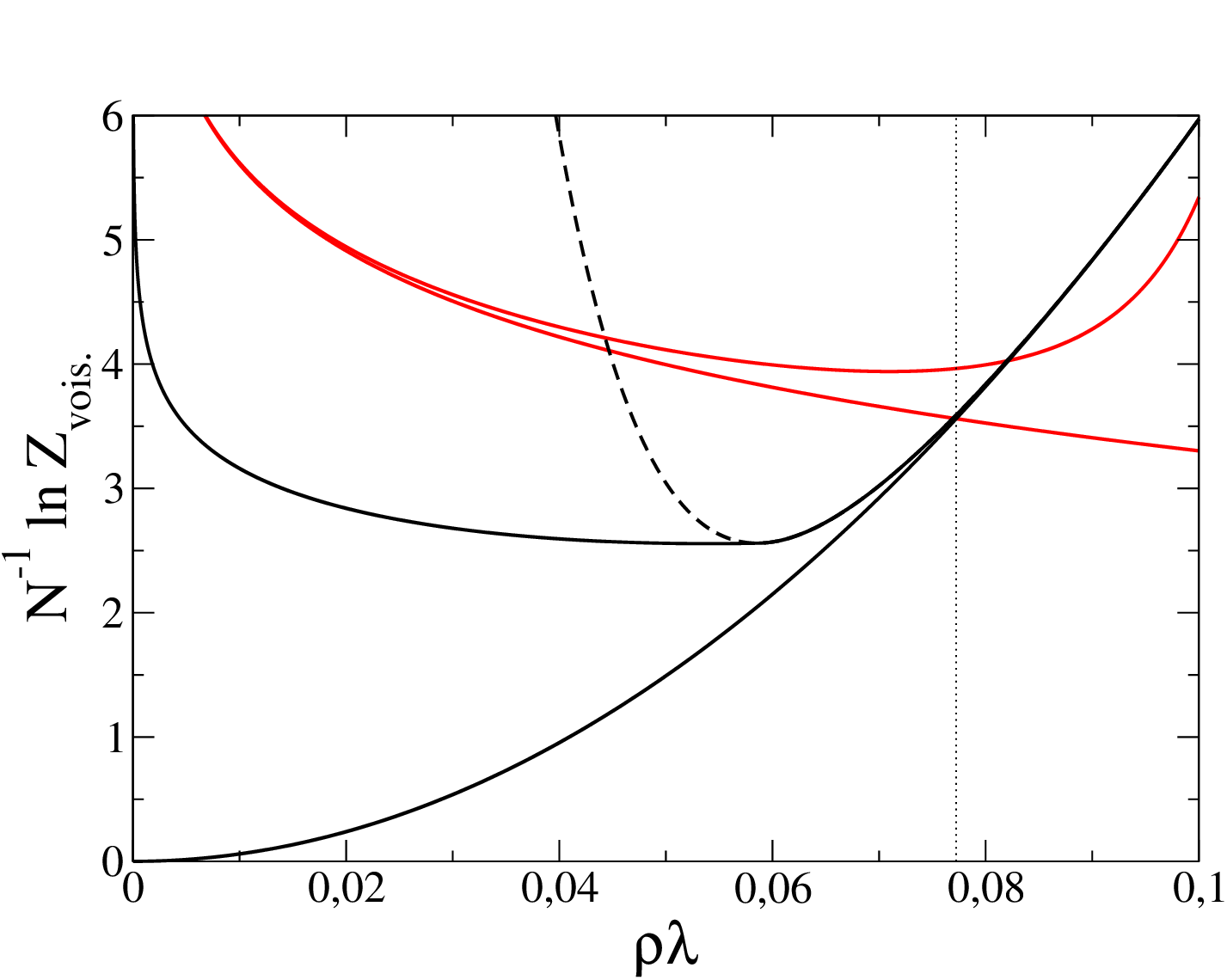}}
\caption{Bracketings of the partition functions $Z_{\rm vois. at.}^{(n\leq N/2)}$ and $Z_{\mathrm{vois.} N\mbox{\tiny -m\`ere}}^{(k\leq N/2)}$, more precisely of their logarithm per particle $N^{-1}\ln Z_{\rm vois.}$, as functions of $\rho \lambda$  for $L/\ell=150$  in the
thermodynamic limit $N\to +\infty$. For the gaseous phase, the bracketing corresponds to the lower bound and the upper bound in (\ref{eq:majat})
(red solid lines). For the liquid phase, the bracketing corresponds to the lower bound and the upper bound in (\ref{eq:maj_nouv})
(black solid lines), or to the upper bound in (\ref{eq:majliq}) (black dashed line). The value of $\rho \lambda$ at the zeroth
order estimate $T_c^{(0)}$ of the transition temperature is shown by the vertical dotted line, at the intersection of the lower red and black
solid lines. The improved upper bound (\ref{eq:maj_nouv}), contrarily to the one of (\ref{eq:majliq}), excludes the existence
of a phase transition at a temperature arbitrarily higher than $T_c^{(0)}$.}
\label{fig:appli}
\end{figure}

In the case of the maximally extended neighbourhoods of the atomic phase  ($n_{\rm max}=N/2$) 
and of the $N$-mer ($k_{\rm max}=N/2$), we show, in figure \ref{fig:appli}, the lower bound and the upper bound
of $N^{-1} \ln Z_{\rm vois.}$, where $Z_{\rm vois.}$ is the partition function of each neighbourhood, as deduced from
the bracketings (\ref{eq:majat}), (\ref{eq:majliq}) and (\ref{eq:maj_nouv}), as a function of $\rho\lambda$ and for $L/\ell=150$.
Contrarily to (\ref{eq:majliq}), the improved bracketing allows to conclude without ambiguity that the liquid-gas transition takes place
at a critical temperature close to $T_c^{(0)}$, even if one takes into account arbitrarily small values of $\rho \lambda$, keeping
in mind that the thermodynamically favoured phase is, at fixed temperature, the one with the largest partition function (that is, with
the smallest free energy).

\end{document}